%% file: ms.tex
\documentclass[twocolumn]{aastex61}

\usepackage{amsmath}
\usepackage{amssymb}

\newcommand{\sqdeg}{deg$^2$ }
\newcommand{\sqdegnospace}{deg$^2$}
\newcommand{\lcdm}{\ensuremath{\Lambda \mathrm{CDM}}}
\newcommand{\omb}{\ensuremath{\Omega_b h^2}}
\newcommand{\omc}{\ensuremath{\Omega_c h^2}}

\newcommand{\sigom}{\ensuremath{\sigma_8 \Omega_{\rm m}^{0.25}}}

\newcommand{\alens}{\ensuremath{ A_{L}}}
\newcommand{\aphiphi}{\ensuremath{ A^{\phi \phi}}}
\newcommand{\mnu}{\ensuremath{ M_{\nu}}}

\newcommand{\planck}{{\textit{Planck}}}

\newcommand{\sptplanck}{{\textsc{SPT} + \textit{Planck}}}
\newcommand{\sptsz}{{\textsc{SPT-SZ}}}
\newcommand{\sptpol}{{\textsc{SPTpol}}}
\newcommand{\bk}{{\textsc{BICEP2}+\textsc{Keck Array}}}

\newcommand{\actpol}{{\textsc{ACTPol}}}
\newcommand{\polarbear}{{\textsc{POLARBEAR}}}

\newcommand*{\wpair}[2]{\genfrac{}{}{0pt}{}{#1}{#2}}

\newcommand{\bi}{\begin{itemize}}
\newcommand{\ei}{\end{itemize}}

\def\apjl{ApJL}

\def\aap{A\&A}

\def\apjs{ApJ Supp.}

\begin{document}

\title{Constraints on Cosmological Parameters from the Angular Power Spectrum of a 
Combined 2500 \sqdeg SPT-SZ and \planck\ Gravitational Lensing Map}

\input{spt_authorlist_v1.tex}

\shorttitle{SPT Lensing Parameter Constraints}
\shortauthors{Simard, Omori, and the SPT collaboration}
\correspondingauthor{Gilbert Holder}
\email{gholder@illinois.edu}

\begin{abstract}
We report constraints on cosmological parameters from the angular power spectrum
of a cosmic microwave background (CMB) gravitational lensing potential 
map created using temperature data from 2500 \sqdeg of South Pole Telescope (SPT) data supplemented with data from \planck\ in the same sky region,
with the statistical power in the combined map primarily from the SPT data.
We fit the corresponding lensing angular power spectrum to a
model including cold dark matter and a cosmological constant 
(\lcdm), and to models with single-parameter
extensions to \lcdm. 
We find constraints that are comparable to and consistent with 
constraints found using the full-sky \planck\ CMB lensing data.
Specifically, we find $\sigom=0.598 \pm 0.024$ from the lensing data alone with relatively
weak priors placed on the other \lcdm\ parameters.
In combination with
primary CMB data from \planck, we explore single-parameter extensions to the \lcdm\ model.
We find $\Omega_k = -0.012^{+0.021}_{-0.023}$ or $\mnu < 0.70$~eV both at 95\% confidence, all in good
agreement with results from \planck\ that include the lensing potential as
measured by \planck\ over the full sky. 
We include two independent free parameters  
that scale the effect of lensing on the CMB: \alens, 
which scales the lensing power spectrum in both the 
lens reconstruction power and in the smearing of the acoustic peaks,
and \aphiphi, which scales only the amplitude of the CMB lensing
reconstruction power spectrum. We find $\aphiphi \times \alens =1.01 \pm 0.08$ for the lensing map made from combined SPT and \planck\ temperature data, 
indicating that the
amount of lensing is in excellent agreement with what is expected from the observed
CMB angular power spectrum when not including
the information from smearing of the acoustic peaks. 
\end{abstract}


\section{Introduction}

Gravitational lensing of the cosmic microwave background (CMB)
has emerged as a useful cosmological tool. 
CMB lensing, which probes all structure 
along a given line of sight, 
provides complementary information to the primary CMB fluctuations
which measure structure at $z\sim 1100$. 
The sensitivity of CMB lensing peaks at intermediate redshifts 
($z\sim 3$), making it complementary to large-scale structure surveys, 
the sensitivity of which typically peaks at lower redshifts, and with very
different sources of possible systematic errors. 
Lensing of the CMB was first detected a decade ago
\citep{smith07};
high signal-to-noise detections have now been achieved by many 
experiments \citep{das11,vanengelen12,planck13-17, polarbear14c,bicep2keck16}. 
For a review of CMB lensing, see \citet{challinor05}.

The fluctuations in the CMB lensing potential form a 
nearly Gaussian projected field on the sky, with statistical properties determined
by the geometry and the history of structure formation in the
universe. Because the field is nearly Gaussian, essentially all
the information is encoded in the angular power spectrum. 
The most precise CMB lensing power spectrum
measurements to date are from the \planck\ experiment \citep{planck15-15}.

Cosmological parameter fits that include CMB lensing information are
broadly consistent with expectations from the primary CMB measurements
alone \citep{planck15-13}. There are, however, mild but interesting tensions
($\sim 2\sigma$) between constraints on cosmology from \planck\ primary CMB measurements
and other cosmological probes.
Specifically related to lensing, the amplitude of
the matter power spectrum on galaxy scales ($\sigma_8$) inferred from \planck\
primary CMB data is slightly higher than that determined from cosmic shear
measurements \citep{joudaki17,hildebrandt17,troxel18}. 
Further, specifically related to lensing of the CMB,
the amount of CMB lensing inferred from the measured smearing of the
acoustic peaks is higher than the amount of CMB lensing inferred from
the direct measurement of the lensing-induced mode-coupling 
\citep{planck15-13}. 
The amplitude of lensing is expected to be a powerful probe of
neutrino masses \citep{abazajian15b}, so discordance in 
measurements of lensing amplitudes is important for understanding
the utility of these measurements as probes of particle physics. 

\begin{figure*}
\centering
\includegraphics[width=0.9\textwidth]{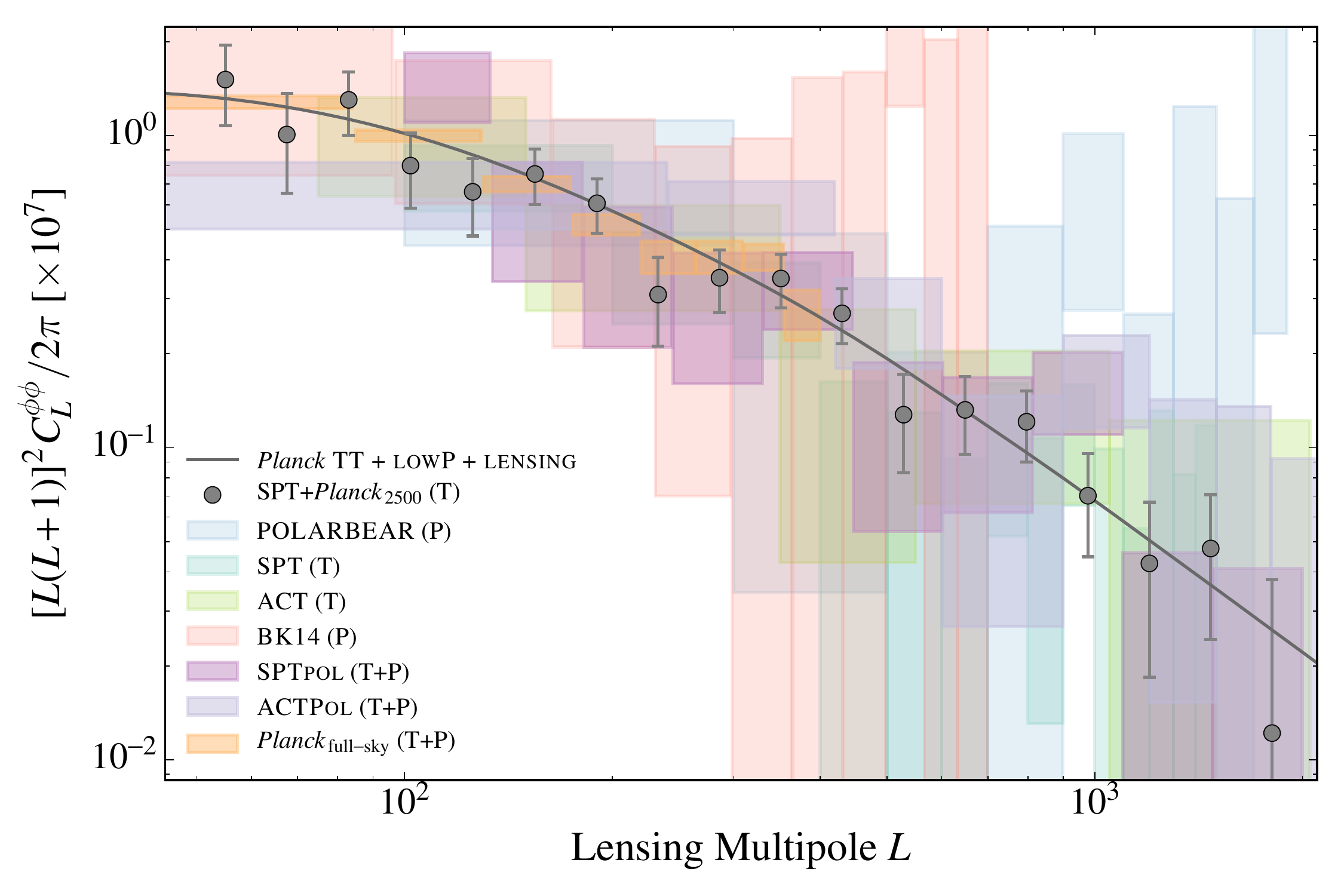}
\caption{  
\sptplanck\ lensing bandpowers from O17 along with earlier lensing estimates from the \sptsz\ survey 
\citep{vanengelen12} and recent lensing bandpowers obtained from temperature and polarization 
measurements from \sptpol\ \citep{story15}. 
Also plotted are the most recent lensing autospectrum measurements from \bk\ \citep{bicep2keck16}, 
\planck\ \citep{planck15-15},  \polarbear\  \citep{polarbear14c} and \actpol\ \citep{sherwin16}, and 
a prediction for the lensing power spectrum using the best-fit cosmological parameters
from the \textsc{\planck\ TT + lowP + lensing} cosmology \citep{planck15-13}.}
\label{fig:bp}
\end{figure*}

This paper is a companion to \citealt{omori17}, referred to as O17 hereafter. 
In that work, we obtained a CMB temperature map by combining 150 GHz SPT 
and 143 GHz \planck\ 
data in the 2500~deg$^2$ SPT-SZ survey region, and we used the resulting temperature
map to produce a map of the projected gravitational lensing potential. 
In this paper, we present a cosmological parameter analysis of the
CMB lensing power spectrum derived in O17. The power spectrum from O17
is shown in Figure~\ref{fig:bp}, along with other recent measurements, including
the full-sky \planck\ lensing power spectrum.

This work is divided as follows: in \S\ref{sec:theory} we 
review gravitational lensing 
of the CMB and reconstruction of the lensing potential; 
in \S\ref{sec:data} we describe the CMB temperature data and 
simulations used for the O17 analysis and for this work; 
in \S\ref{sec:likelihood} we describe how the lensing likelihood is constructed, including
linear corrections for the unknown true CMB and lensing 
potential power spectra; in \S\ref{sec:constraints}
we present the primary result of this paper: constraints on cosmological parameters; 
we close with a discussion.

	Throughout this work, we use the \textsc{\planck\ TT + lowP + lensing} cosmology\footnote{base\_plikHM\_TT\_lowTEB\_lensing} \citep{planck15-13} as a fiducial model. This fiducial cosmology is used for generating the simulated data necessary for the lensing reconstruction. 
All CMB temperature and lensing potential power spectra used in the present analysis have been computed with the \textsc{CAMB} Boltzmann code\footnote{\url{http://camb.info} - May 2016 version} \citep{lewis00}. 

\section{Lensing Reconstruction Framework}
\label{sec:theory}

In this section, we build the theoretical framework for the lensing 
likelihood, presenting selected elements from the lensing reconstruction 
pipeline.  A more complete description of the procedure can be found in O17. 

\subsection{Lensing of CMB Temperature Fluctuations}

Gravitational lensing remaps CMB fluctuations in position space 
\citep{lewis06}:
\begin{equation} \label{eq:remap}
T^{\text{L}}(\mathbf{\hat{n}}) \, = T^{\text{U}}( \mathbf{\hat{n}} +  \nabla\phi(\mathbf{\hat{n}}) ),
\end{equation}
where $\phi( \mathbf{\hat{n}} )$ is the projected gravitational
lensing potential and L and U refer to the lensed and unlensed temperature fields respectively. 
To gain intuition, Equation~\ref{eq:remap} can be Taylor expanded as
\begin{equation} \label{eq:lens_exp}
T^{\text{L}} ( \mathbf{\hat{n}} ) = T^{\text{U}}( \mathbf{\hat{n}} ) + \nabla T^{\text{U}} \cdot \nabla \phi (\mathbf{\hat{n}} ) + \ldots
\end{equation}
From the second term, it can be seen that the observed lensed temperature has a component that is the gradient of the unlensed field 
modulated by the lensing deflection $\nabla \phi$.
If we transform to harmonic space, Equation \ref{eq:lens_exp} would have the
second term on the right hand side written as a weighted convolution of the temperature field and
the lensing potential, where the harmonic transform 
for any particular mode for the lensed field could involve a sum over all of the modes of the unlensed field. 
Lensing thus introduces non-zero off-diagonal elements in the covariance of observed temperature fields in
harmonic space \citep{okamoto03}: 
\begin{eqnarray}
\label{eq:mode_coupling}
\Delta \left\langle  T_{\ell_1 m_1} T_{\ell_2 m_2}  \right \rangle \\
= && \sum_{LM} (-1)^M \left (  \wpair{\ell_1}{m_1}  \wpair{\ell_2}{m_2} \wpair{L}{-M}  \right ) W_{\ell_1 \ell_2 L}^{\phi}\,\, \phi_{LM}, \nonumber
\end{eqnarray}
where $T_{\ell m}$ are the spherical harmonic expansion coefficients of the temperature fields and $\phi_{LM}$ the coefficients of the projected lensing potential. The weight 
\begin{eqnarray}
\label{eq:weight_function}
W_{\ell_1 \ell_2 L}^{\phi} = && - \sqrt{\frac{(2\ell_1 + 1)(2\ell_2 + 1)(2L + 1)}{4\pi}} \\
&& \times C_{\ell_1}^{TT}\left( \frac{1 + (-1)^{\ell_1 + \ell_2 + L}}{2} \right)
\left (  \wpair{\ell_1}{1}  \wpair{\ell_2}{0} \wpair{L}{-1}  \right ) \nonumber \\
&& \times \sqrt{L(L+1)\ell_1(\ell_1 + 1)} + (\ell_1 \leftrightarrow \ell_2)\nonumber 
\end{eqnarray}
characterizes the mode coupling induced by
lensing (i.e., the effect of the convolution in Equation~\ref{eq:lens_exp}). 

\subsection{Lensing Map Reconstruction}

The lensing potential can be estimated from observed CMB maps by measuring the 
lensing-induced mode coupling of Equation \ref{eq:mode_coupling} between pairs of modes
in the observed temperature field (\citealt{zaldarriaga99}; \citealt{hu02a}).  
In general, it is best to use pairs in harmonic space that have good signal-to-noise 
for measuring lensing. For this purpose, it is useful to work with a filtered
map: $\bar{T}_{\ell m} \equiv F_{\ell m} T_{\ell m}$, with the filter 
$F_{\ell m}\equiv (C_\ell + N_{\ell m})^{-1}$ for a given CMB power 
spectrum $C_\ell$ and an anisotropic ($m$-dependent) 
noise power spectrum $N_{\ell m}$.

A formally optimal estimator (at first order) which maximizes signal to noise in the estimated
lensing potential \citep{hu02a} is
\begin{eqnarray}
\label{eq:phi_bar}
\bar{\phi}_{LM} \\ 
= && \frac{(-1)^M}{2} \sum_{ \substack{ \ell_1, m_1 \\ \ell_2, m_2 } } \left(  \wpair{\ell_1}{m_1}  \wpair{\ell_2}{m_2} \wpair{L}{-M}  \right ) W_{\ell_1 \ell_2 L}^{\phi} \,\, \overline{T}_{\ell_1 m_1} \overline{T}_{\ell_2 m_2}. \nonumber
\end{eqnarray}

We use Equation \ref{eq:phi_bar} as our $\phi$ estimator for this analysis. 
There are other choices (e.g., \citealt{namikawa13})  
for how to weight the mode pairs which sacrifice some signal-to-noise
but reduce foreground contamination.
Lensing reconstruction is 
done with the \textsc{quicklens} code.\footnote{\url{http://github.com/dhanson/quicklens}}

The relationship between the filtered estimate of the lensing potential resulting
from Equation \ref{eq:phi_bar} and the true potential can be written as
\begin{equation}
\bar{\phi}_{L M } \equiv \mathcal{R}_{L M}^{\phi}\phi_{L M} \ ,
\end{equation}
defining a response function $\mathcal{R}_{L M}$ that in general
depends on both $L$ and $M$. As outlined in O17, this
response function has been calibrated using simulations. We estimate the response function
by measuring the cross-spectrum of simulated lensing potential outputs with the input 
lensing potential maps and normalizing by the autospectrum of the inputs. 

The true amplitude of mode coupling in the CMB temperature field induced by lensing
is sensitive to the true (unknown) temperature
power spectrum, as can be seen in Equations~\ref{eq:mode_coupling} and \ref{eq:weight_function}.
What is measured in the data is some amount of mode coupling; to turn this into
an estimate of the amplitude of the lensing potential, an assumption is made
about the typical amplitudes of the modes being coupled. 
The response function thus depends on the assumed cosmological parameters. 
To explore
this cosmological dependence, we use an isotropic approximation to the full anisotropic
response function and its dependence on cosmology.
In the case where both the signal and noise are
isotropic (i.e., the CMB signal and noise only depend statistically on $\ell$ and not $m$), the
response function can be written as
\begin{equation}
\mathcal{R}_{L}^{\phi} = \frac{1}{2 L + 1}  \sum_{\ell_1, \ell_2} W_{\ell_1 \ell_2 L}^{\phi,t} 
W_{\ell_1 \ell_2 L}^{\phi,f} F_{ \ell_1} F_{ \ell_2}  \  , 
\label{eqn:isotropic_R}
\end{equation} 
where we have indicated extra superscripts on the weight functions for either the 
true amount of mode coupling ($t$) or the assumed amount for our fiducial
cosmology ($f$). The filters $F_\ell$ are calculated for the fiducial cosmology.
We use Equation \ref{eqn:isotropic_R} and its dependence on cosmology 
to determine the cosmology-dependent corrections to the simulation-based response function. 

The survey mask, point source mask, and spatially varying noise all violate statistical stationarity in the data, and consequently they introduce mode coupling that can bias the lensing reconstruction.
The result is that the lensing reconstruction has
a non-zero mean signal---even in the absence of true lensing signal---that 
depends on the survey geometry, mask, and noise properties. 
This mean field $\bar{\phi}_{L M}^{MF}$ is calculated using simulations and removed. 

After removing the mean field and correcting
for the response function, the final estimate of the lensing potential is
\begin{equation}
\hat{\phi}_{L M} = \frac{\bar{\phi}_{L M}-\bar{\phi}_{LM}^{\rm MF}}{\mathcal{R}_{LM}^\phi} \ .
\end{equation}

\subsection{Lensing Autospectrum Estimation}
\label{subsec:lens_clpp}

To estimate the angular power spectrum of the CMB lensing map obtained in the previous
section, we multiply the estimate $\hat{\phi}$ by the survey mask (including point source
and galaxy cluster masking) and use
{\tt PolSpice}\footnote{\url{http://www2.iap.fr/users/hivon/software/PolSpice}} \citep{szapudi01,chon04}
to compute the spectrum of the masked map.

The resulting power spectrum is a biased estimate of the true lensing
power spectrum. Known sources of bias include a straightforward noise bias,
$N_L^{(0)}$, that comes from taking an autospectrum of data with noise in it
(where ``noise'' here includes the Gaussian part of the CMB temperature field and any other sky signal), 
and a bias that arises from ambiguity in exactly which lensing modes are being measured
in the power spectrum, $N_L^{(1)}$ \citep{kesden03}. 
The superscript denotes the order of the lensing power spectrum involved: 
$N_L^{(0)}$ is independent of the true lensing power and only depends on the
instrument noise and sky power, while $N_L^{(1)}$ has a linear dependence
on the lensing power. As detailed in O17, we calculate these biases using
simulations and subtract them from the measured power spectrum:
\begin{equation}
C_L^{\hat{\phi}\hat{\phi}} = \hat{C}_L^{\hat{\phi}\hat{\phi}} - N^{(0)}_{L} - N^{(1)}_{L}  \ .
\end{equation}
We use a realization-dependent $N_L^{(0)}$ estimate that takes into account 
the power in the particular realization but does not depend on the 
assumed cosmology
\citep{namikawa13}.

The $N_L^{(1)}$ bias depends linearly on the lensing power and
will therefore depend on cosmological parameters.
In the flat-sky limit
\citep{das11,kesden03, planck13-17} and assuming isotropic noise and
filtering, the bias is 
\begin{eqnarray}
\label{eqn:isotropic_N1}
N^{(1)}_{L} = && \frac{1}{\mathcal{R}_{L}^{\phi} \mathcal{R}_{L}^{\phi}}
\int \frac{d^2 \boldsymbol{\ell_1} }{(2\pi)^2}  \int \frac{d^2 \boldsymbol{\ell_3} }{(2\pi)^2}  \\ \nonumber
&& \times \, F_{ \ell_1} \, F_{ \ell_2} \, F_{ \ell_3} \, F_{ \ell_4} \,\, W^{\phi, f}({\boldsymbol{\ell}_1, \boldsymbol{\ell}_2})
\, W^{\phi, f}({\boldsymbol{\ell}_3, \boldsymbol{\ell}_4}) \\ \nonumber
&& \times \Big[ C^{\phi\phi}_{ | \boldsymbol{\ell_1} - \boldsymbol{\ell_3} |}
W^{\phi, t}({-\boldsymbol{\ell}_1, \boldsymbol{\ell}_3}) \, W^{\phi, t}({-\boldsymbol{\ell}_2, \boldsymbol{\ell}_4}) \\ \nonumber
&& + \,\,\, C^{\phi\phi}_{| \boldsymbol{\ell_1} - \boldsymbol{\ell_4} |}
W^{\phi, t}({-\boldsymbol{\ell}_1, \boldsymbol{\ell}_4}) \, W^{\phi, t}({-\boldsymbol{\ell}_2, \boldsymbol{\ell}_3}) \Big],
\end{eqnarray}
where the weight $W^{\phi}({\boldsymbol{\ell}_1, \boldsymbol{\ell}_2})$ is the flat-sky version of Equation \ref{eq:weight_function}. 

There is a dependence on both the true CMB power (just as for $\mathcal{R}^\phi_L$)
and the lensing power. To explore this cosmological dependence (below), 
we will use Equation \ref{eqn:isotropic_N1} to determine the cosmology-dependent
corrections to the $N_L^{(1)}$ that is derived from simulations.

The next-order $N_L^{(2)}$ bias is largely removed by using the lensed 
theory temperature power
spectrum rather than the unlensed spectrum when constructing the lensing
estimator \citep{hanson11}.  
There are other biases, such as the $N_L^{(3/2)}$ bias \citep{bohm16}, 
which are small at the precision of the current work, and will be neglected. 

We estimate uncertainties on the lensing power spectrum by averaging over $N_s=198$ simulations:
\begin{equation}
(\Delta C_L^{\hat{\phi}\hat{\phi}})^2 = \frac{1}{N_{s}-1} \sum_{i=1}^{N_{s}}  (C_{L;i}^{\hat{\phi}\hat{\phi}} - \langle C_{L}^{\hat{\phi}\hat{\phi}} \rangle_{N_{s}} )^2 \ .
\end{equation}
This procedure could be used to generate a full covariance matrix, but
for this analysis we assume that uncertainties are uncorrelated between
bins. This is expected for the relatively large bins that we use and the 
realization-dependent removal of the $N_L^{(0)}$ bias that strongly reduces
the off-diagonal elements of the covariance matrix \citep{schmittfull13}.

\section{Lensing Data}
\label{sec:data}

The binned CMB lensing angular power spectrum  
(or ``lensing bandpowers'') $\hat{C}_{L_{\text{b}}}^{\phi\phi}$ 
computed in O17, using the methods
described in that work and summarized in the previous section, is shown in 
Figure~\ref{fig:bp} (along with other recent measurements from the literature),
and the bin ranges and bandpower values and uncertainties
are listed in Table~\ref{tab:bp}.\footnote{\url{https://pole.uchicago.edu/public/data/simard18}} 
We will hereafter refer to this as the ``\sptplanck'' lensing measurement.

\begin{figure}
\includegraphics[width=\columnwidth]{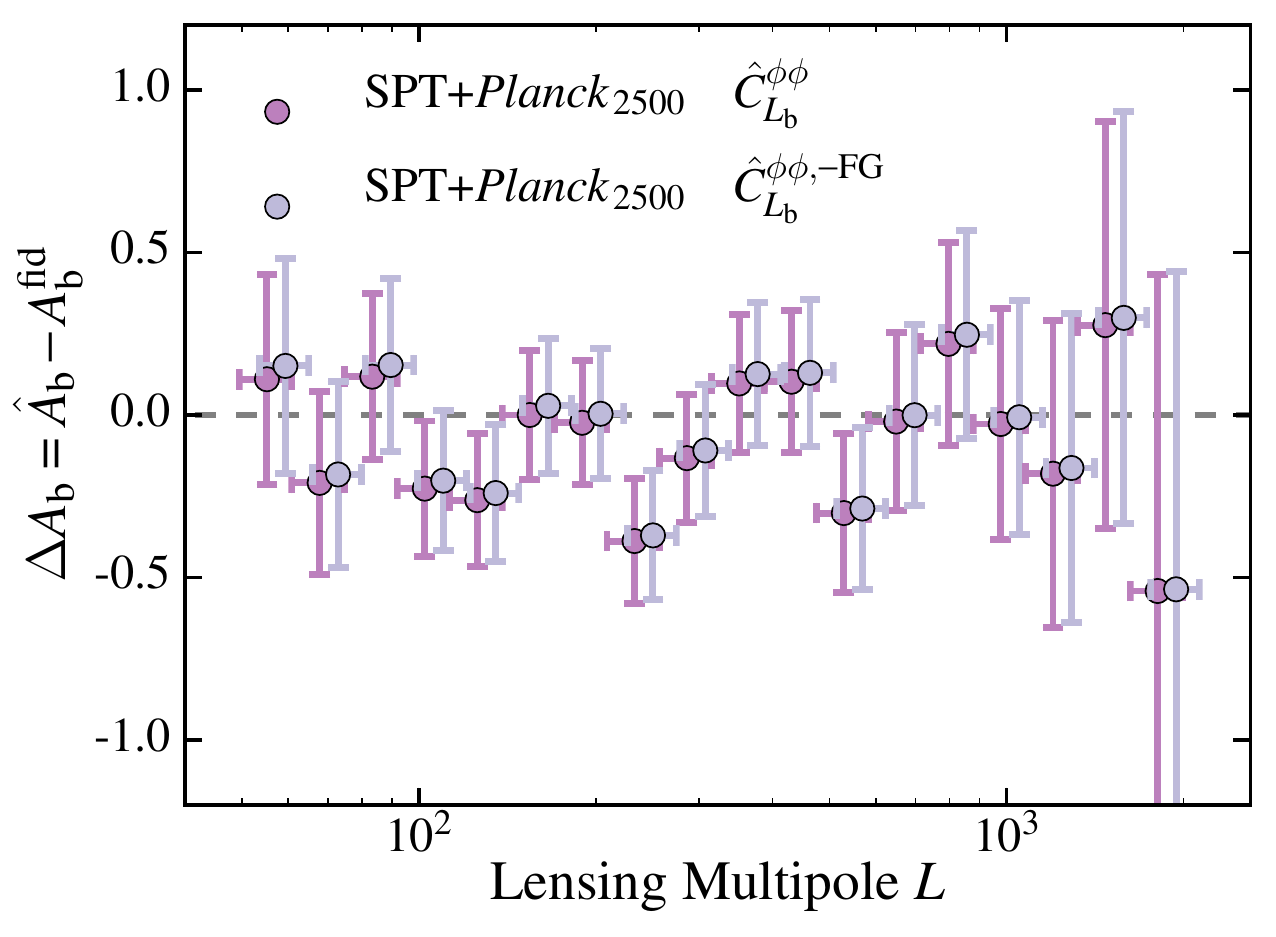}
\caption{Ratio of lensing bandpowers to lensing power spectrum predicted
for the best-fit \planck\ 2015 cosmological parameters. Shown are both the raw
lensing bandpowers and the results after subtracting the best estimate
of foreground contamination.}
\label{fig:ratio}
\end{figure}

\begin{deluxetable}{cc}
\tablecolumns{2}
\tablecaption{Foreground removed lensing bandpowers used in this analysis. \label{tab:bp}}
\tablehead{   
  \colhead{$L$} & \colhead{$[L_{\text{b}}(L_{\text{b}} + 1)]^2 \, \hat{C}_{L_{\text{b}}}^{\phi\phi} /2\pi \, [\times 10^7]$} 
  }
\startdata
50-60 & 1.51 $\pm$ 0.44 \\
61-74 & 1.01 $\pm$ 0.35\\
75-91 & 1.30 $\pm$ 0.30\\
92-112 & 0.80 $\pm$  0.22\\
113-138 & 0.66 $\pm$  0.18 \\
139-170 & 0.75 $\pm$  0.15 \\ 
171-209 & 0.61 $\pm$  0.12 \\ 
210-256 & 0.309 $\pm$  0.098 \\  
257-315 & 0.350 $\pm$  0.080 \\  
316-386 & 0.348 $\pm$  0.068 \\  
387-474 & 0.269 $\pm$  0.054\\  
475-582 & 0.128 $\pm$  0.045 \\ 
583-715 & 0.132 $\pm$ 0.037 \\
716-877 &  0.121 $\pm$ 0.031 \\
878-1077 &  0.070 $\pm$ 0.025 \\ 
1078-1322 &  0.043 $\pm$ 0.024 \\
1323-1622 & 0.048 $\pm$ 0.023 \\
1623-1991 & 0.012 $\pm$ 0.026  \\
\enddata
\end{deluxetable}

The higher angular resolution of SPT greatly increases the lensing
signal-to-noise per pixel over \planck\ from the larger number
of available small-scale modes which can be used for measuring the 
lensing-induced mode coupling.
Combining the \planck\ and SPT temperature 
maps strongly reduces the uncertainties in particular on small scales
(higher $L$) as compared to using only the SPT data. This happens because the lensing map only uses modes in the
temperature map extending to $\ell=3000$, to minimize possible foreground
contamination. The high $L$ lensing modes require probing correlations in
the temperature angular modes that are widely separated in harmonic space.
By using the \planck\ data to recover the low-$\ell$ modes, there is an
increased number of large-separation mode pairs.

As shown in O17, the 
\sptplanck\ measurements over the 2500 \sqdeg SPT-SZ survey area
are more precise than the \planck-only full-sky constraints 
for $L\gtrsim 1000$. From the relative sky coverage, the \planck-only 
uncertainties using only the SPT region would be more than three times
larger than the \planck-only full-sky constraints. The combined
\sptplanck\ measurements are thus nearly statistically independent, adding
substantial new information.

Small-scale lensing measurements are most susceptible to foreground contamination,
as shown in \citet{vanengelen14}. In that work, it was found that 
foreground contamination increased dramatically beyond $L\sim 2000$ for CMB map
filtering choices similar to those adopted in O17.
For the cosmological parameter estimation in this work, we therefore 
use the \sptplanck\ lensing measurements only below $L=2000$. 

A comparison of the O17 bandpowers with the prediction from the best-fit \planck\ cosmology is shown in Figure
\ref{fig:ratio}. 
The ratio is shown with and without a correction for foreground contamination, based
on \citet{vanengelen14}.
The estimated contamination is small, but not completely negligible.
The O17 bandpowers are consistent with
expectations from \planck, 
with O17 finding a relative amplitude of 
$0.95 \pm 0.06$ for the best-fit \textsc{\planck\ TT + lowP + lensing} cosmology.

In the likelihood analysis described in the following section, the theory model includes this mean foreground
contamination, as well as a term in the covariance
to account for uncertainty in the foreground level.

\section{Lensing Likelihood}
\label{sec:likelihood}

In this section, we describe how we obtain the lensing likelihood function for the \sptplanck\ lensing data
as a function of cosmological parameters, 
$\ln \mathcal{L}(\mathbf{\Theta})$: 
\begin{eqnarray}
-2 \ln \mathcal{L}( \boldsymbol{\Theta}) \\
 =  \sum_{i,j} &&
\Big[\hat{C}_{L^i_{\text{b}}}^{\phi\phi} - C_{L^i_{\text{b}}}^{\phi\phi, \mathrm{th}}(\boldsymbol{\Theta}) \Big]
\mathbb{C}^{-1}_{L^i_{\text{b}} L^j_{\text{b}}}
\Big[\hat{C}_{L^j_{\text{b}}}^{\phi\phi} - C_{L^j_{\text{b}}}^{\phi\phi, \mathrm{th}}(\boldsymbol{\Theta}) \Big]. \nonumber
\end{eqnarray}
We make the approximation that the reconstructed lensing bandpowers $\hat{C}_{L_{\text{b}}}^{\phi\phi}$ 
are Gaussian-distributed and uncorrelated between bins, 
but that there is correlation between
bins coming from the uncertainty in the foreground subtraction. 
We assume that the uncertainty in the residual foreground as reported in 
\citet{vanengelen14} is completely correlated between bins, leading to off-diagonal
terms in the covariance matrix.
The $C_{L_{\text{b}}}^{\phi \phi, \mathrm{th}}(\mathbf{\Theta})$ bandpowers 
correspond to the binned theory lensing power spectrum at a given 
cosmology $\mathbf{\Theta}$, with the foreground template added. 

\subsection{Linear Corrections}
\label{subsec:lincorr}

The choice of cosmological model affects the computation of the estimated 
lensing bandpowers through the calculation of the response function 
and through the calculation of the 
$N_L^{(1)}$ bias term.
These effects need to be included in the likelihood analysis.

The response function $\mathcal{R}^{\phi}_{LM}$ and $N_L^{(1)}$ correction 
for the fiducial cosmology are obtained using simulations and calculated
using two-dimensional, anisotropic weighting.
To calculate the cosmological corrections to these terms, we use isotropic approximations
to both the response function and the $N_L^{(1)}$ bias
(see Equations \ref{eqn:isotropic_R} and \ref{eqn:isotropic_N1}). 
Within the range of allowed parameters, the cosmological corrections
are relatively small, and we expect the error on these corrections from 
using the isotropic approximation to be negligible.

At a given point in parameter space, we apply the cosmology-dependent 
response function and $N_L^{(1)}$ corrections to the theory spectrum \citep{planck15-15}:
\begin{equation}
C_{L}^{\phi\phi,\text{th}} = \frac{(\mathcal{R}_{L}^{\phi})^2\big|_{\Theta} \hfill }{(\mathcal{R}_{L}^{\phi})^2\big|_{f}} C_{L}^{\phi\phi, \text{th}}\big|_{\Theta} + N^{(1)}_{L}\big|_{\Theta} - N^{(1)}_{L}\big|_{f}.
\end{equation}
To obtain these corrections, we use a linear approximation, Taylor-expanding
around the response function or $N_L^{(1)}$ bias calculated for the fiducial 
cosmology. For a temperature or lensing power spectrum that
differs by $\Delta$ from the fiducial spectrum, we obtain:
\begin{equation}
\Delta (\mathcal{R}_{L}^{\phi})^2  \big|_{\Theta} \simeq  M^{(R)}_{L,\ell'} 
\big|_{f} \times \Delta C^{TT}_{\ell'} \big|_{\Theta} \ , 
\end{equation}
where $M^{(R)}_{L,\ell'} \equiv \frac{C_{L}^{\phi\phi}}{(\mathcal{R}_{L}^{\phi})^2} \frac{\partial (\mathcal{R}_{L}^{\phi})^2 \hfill}{\partial C_{\ell'}^{TT}}$, and 
\begin{equation}
\Delta N^{(1)}_{L} \big|_{\Theta} \simeq M^{(1)}_{L,L'} \big|_{\text{fid}} \times \Delta C^{\phi \phi}_{L'} \big|_{\Theta}  \  ,
\end{equation}
where $M^{(1)}_{L,L'} \equiv 
\frac{\partial N^{(1)}_{L} }{\partial C_{L'}^{\phi\phi}}$.
The matrices $M$ were calculated using binned versions of the temperature
and lensing power spectra.
In principle there is also a dependence on the temperature power spectrum
in the $N_L^{(1)}$ correction, but that term was found to be negligible.

\section{Constraints on Cosmological Parameters}
\label{sec:constraints}

Sourced mainly by potential wells at intermediate redshifts,
gravitational lensing of the CMB can constrain late-time cosmological parameters affecting the growth of structure and the expansion of the universe, such as 
neutrino masses \citep{smith06b,abazajian15b}, and as a geometrical effect it can
constrain curvature \citep{sherwin11}.
Because of the combined sensitivity to the geometry and the growth of structure, lensing can break degeneracies between cosmological parameters constrained by CMB alone,
including the angular diameter distance degeneracy \citep{stompor99}.

Recent detections of CMB lensing have proven its significance as a cosmological probe, on its own or in combination with CMB temperature and polarization measurements \citep{planck15-15,planck13-17,vanengelen12,das11}.
In the following section, we show the most significant improvements on 
cosmological parameters constraints provided by the \sptplanck\ 
lensing measurements over 2500 \sqdeg
as compared to the full-sky \planck\ primary CMB 
measurements on their own. 

To determine the posterior probability distributions of the cosmological parameters from \sptplanck\ lensing data in combination with CMB data, we use 
Markov Chain Monte Carlo (MCMC) methods 
\citep{christensen01} through the publicly available \textsc{CosmoMC}\footnote{\url{http://cosmologist.info/cosmomc/} - July 2016 version} package \citep{lewis02b}. 

Assuming a spatially flat universe, the properties of a $\Lambda$ cold dark matter (\lcdm) 
model can be represented by the following six parameters, which are the base set of 
parameters to be varied in the chains:
the baryon density \omb,
the cold dark matter density \omc,
the optical depth at reionization $\tau$,
the angular scale of the sound horizon at the surface of last scattering $\theta_s$,
the amplitude $A_s$ and power-law spectral index $n_s$ of primordial scalar perturbations, both taken at a pivot scale of $k=0.05 \, \mathrm{Mpc}^{-1}$ as chosen in the cosmological parameters analysis of \citet{planck13-16}. 
We will often use parameters derived from these six, including the total matter density $\Omega_{\rm m}$.

For constraints based only on lensing, 
the same priors as in \citet{sherwin16} have been applied.
When computing constraints combining CMB lensing measurements with primary CMB 
measurements, the \planck\ TT and lowP likelihoods have been used, 
the latter relying on low $\ell$ CMB temperature and polarization data.

\subsection{\lcdm\ Model}

\begin{figure}[t!]
\includegraphics[width=\columnwidth]{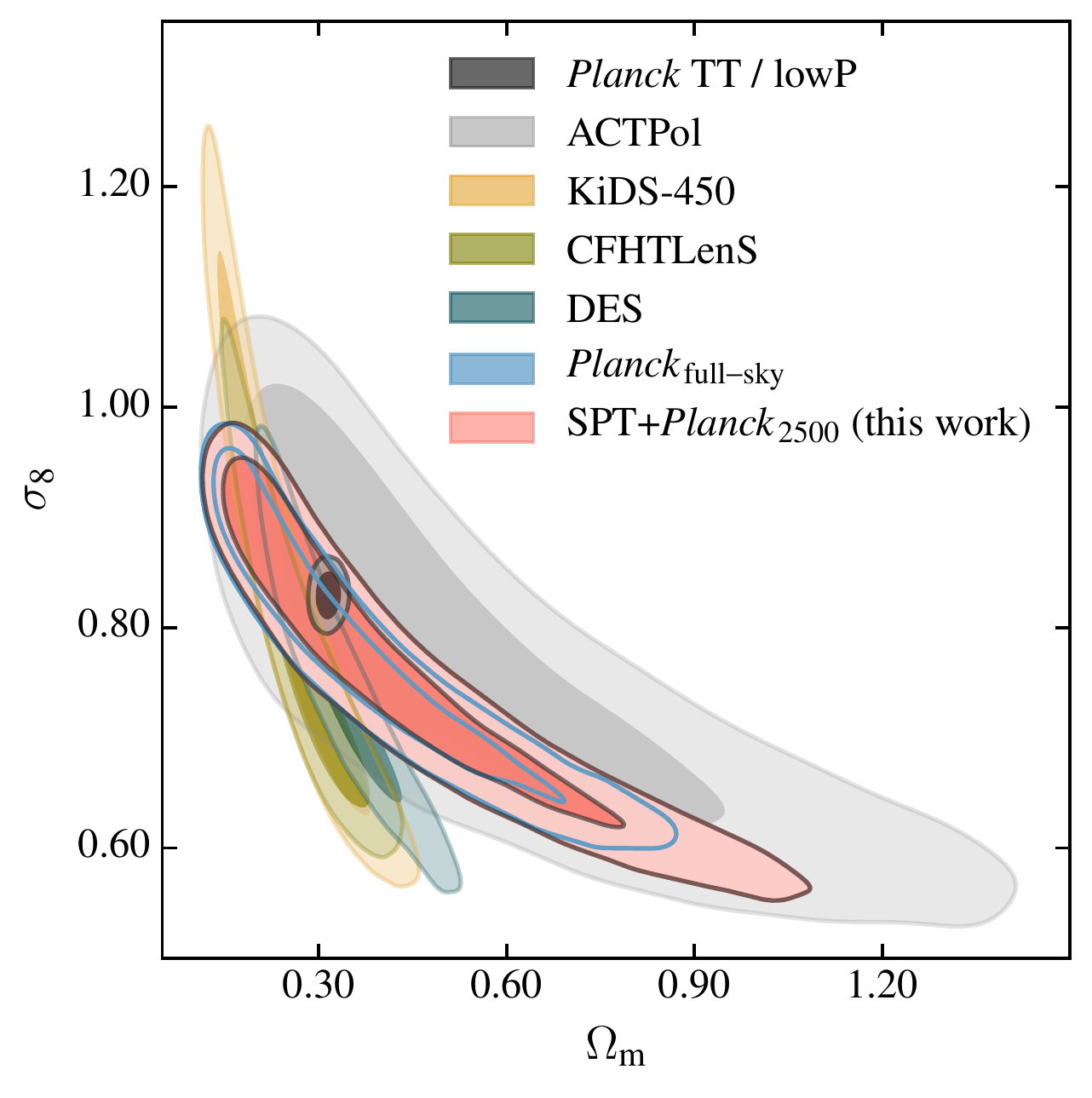}
\caption{Lensing constraints on $\sigma_8$ and $\Omega_{\rm m}$ from optical
surveys (KiDS-450, CFHTLens, DES) and CMB measurements 
(ACTPol, \planck\ full sky, \sptplanck\ 2500 \sqdeg). Also shown
are constraints from the \planck\ primary CMB power spectra. This work is
in good agreement with both CMB and optical surveys.}
\label{fig:lensing_only}
\end{figure}

\begin{figure}
\includegraphics[width=\columnwidth]{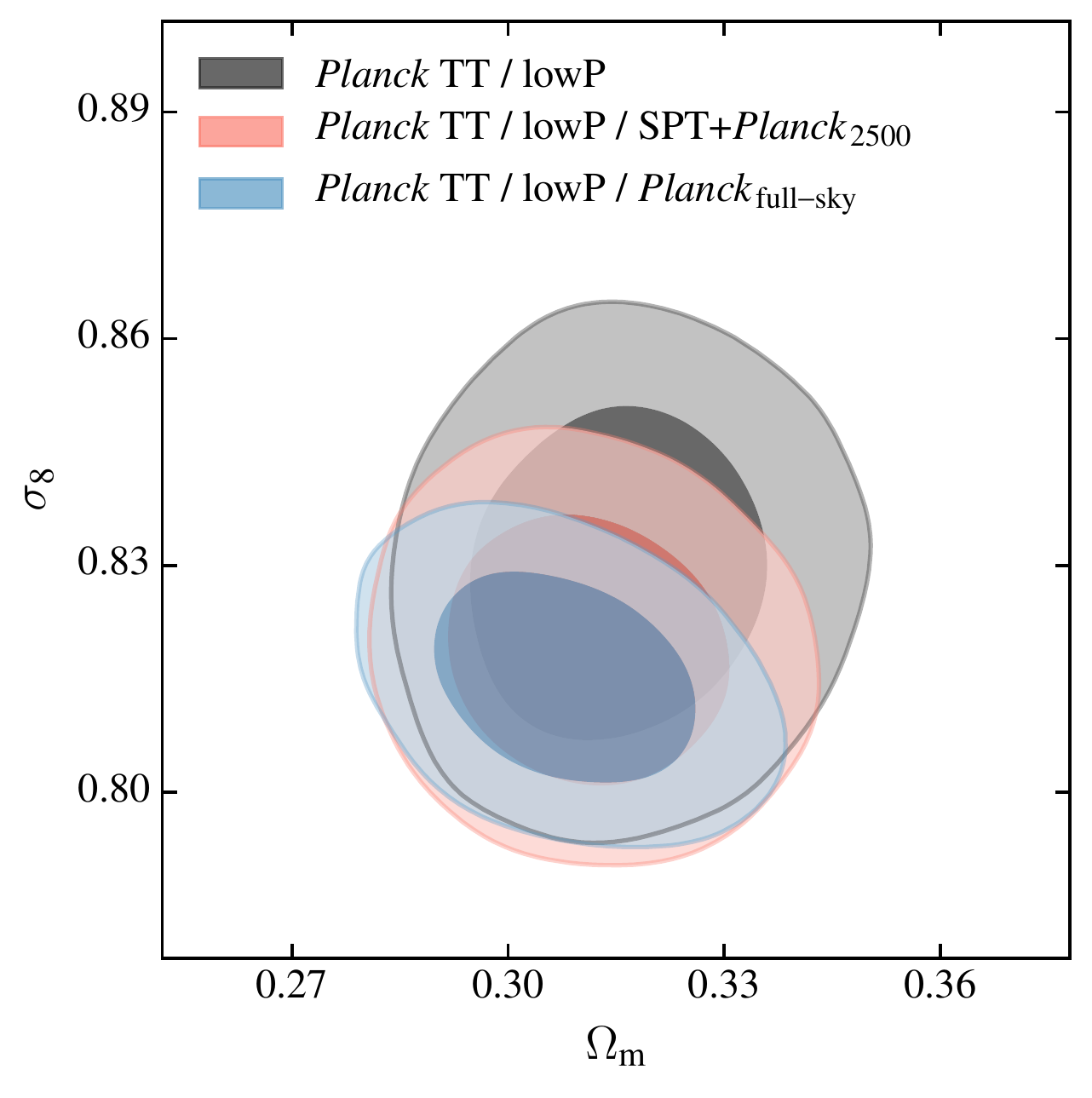}
\caption{Constraints on $\sigma_8$ and $\Omega_{\rm m}$, combining CMB lensing
data with primary CMB constraints. The largest contours show CMB primary
CMB constraints from \planck\, intermediate contours show the impact of
adding the 2500 \sqdeg \sptplanck\ data, the smallest contours show
\planck primary data combined with \planck\ full-sky lensing results. This
work is in excellent agreement with the \planck\ full-sky lensing result.}
\label{fig:sigma8}
\end{figure}

An alternative way to parameterize the amplitude of the matter power spectrum
is $\sigma_8$, the rms mass fluctuation today 
in 8 $h^{-1}$Mpc spheres assuming linear theory. This parameter is 
convenient for comparisons with results from galaxy surveys.

In Figure~\ref{fig:lensing_only}, constraints from lensing experiments,
both CMB lensing \citep{sherwin16,planck15-15} and cosmic shear 
\citep{joudaki17,hildebrandt17,troxel18}, are shown in the 
$\sigma_8-\Omega_{\rm m}$ plane, compared with expectations from
the primary CMB fluctuations as measured by \planck. 
There have been hints of mild tension
between \planck\ CMB power spectrum constraints and probes of low-redshift
structure. The CMB lensing constraints are all highly consistent with each
other, and it can be seen that the constraints from this paper (the
SPT+\planck\ CMB lensing data) overlap with both the low-redshift
probes and the primary CMB estimates, although the primary CMB data are
substantially more precise. In making the CMB-lensing-only constraints,
as was done in \citet{planck15-15} and \citet{sherwin16}, 
the corrections to the response function were held at the best-fit cosmology corresponding to the \planck\ TT and lowP likelihoods\footnote{base\_plikHM\_TT\_lowTEB} \citep{planck15-13}. 
The close agreement between \sptplanck\ and \planck\ is not simply from the combined
\sptplanck\ dataset including data from \planck. The \sptplanck\ data is based
on only $\sim$2500 \sqdeg, and is mainly driven by the SPT data.

Joint constraints on $\Omega_{\rm m}$ and $\sigma_8$
obtained by combining the CMB lensing data with the primary CMB measurements from
\planck\ are shown in Figure~\ref{fig:sigma8}. In general, the
CMB lensing data (either the full-sky \planck\ or 2500 \sqdeg 
\sptplanck) prefer lower values of $\sigma_8$, as could be expected
from Figure~\ref{fig:lensing_only}.
A commonly used parameter for lensing constraints is 
\sigom. For \sptplanck\ 
we find $\sigom=0.598 \pm 0.024$, in excellent
agreement with both the 
value found using the \planck\ full-sky lensing reconstruction, 
$0.591 \pm 0.021$ \citep{planck15-15},
and the estimate
by \actpol\ of $0.643 \pm 0.054$ \citep{sherwin16}.

CMB lensing data are most sensitive to overall shifts in the amplitude of matter fluctuations.
This amplitude can be expressed as the rms deflection angle
$\langle d^2 \rangle^{1/2}$. 
For \sptplanck, we use the samples from the MCMC chains for \lcdm\ to 
determine that this rms deflection angle
is $2.27 \pm 0.16$ arcmin (68\%), in good agreement with the 
extremely precise measurement of the full-sky \planck\ survey of $2.46 \pm 0.06$.

\subsection{Lensing amplitude compared to expectations}

\begin{figure*}
\centering
\includegraphics[width=0.9\textwidth]{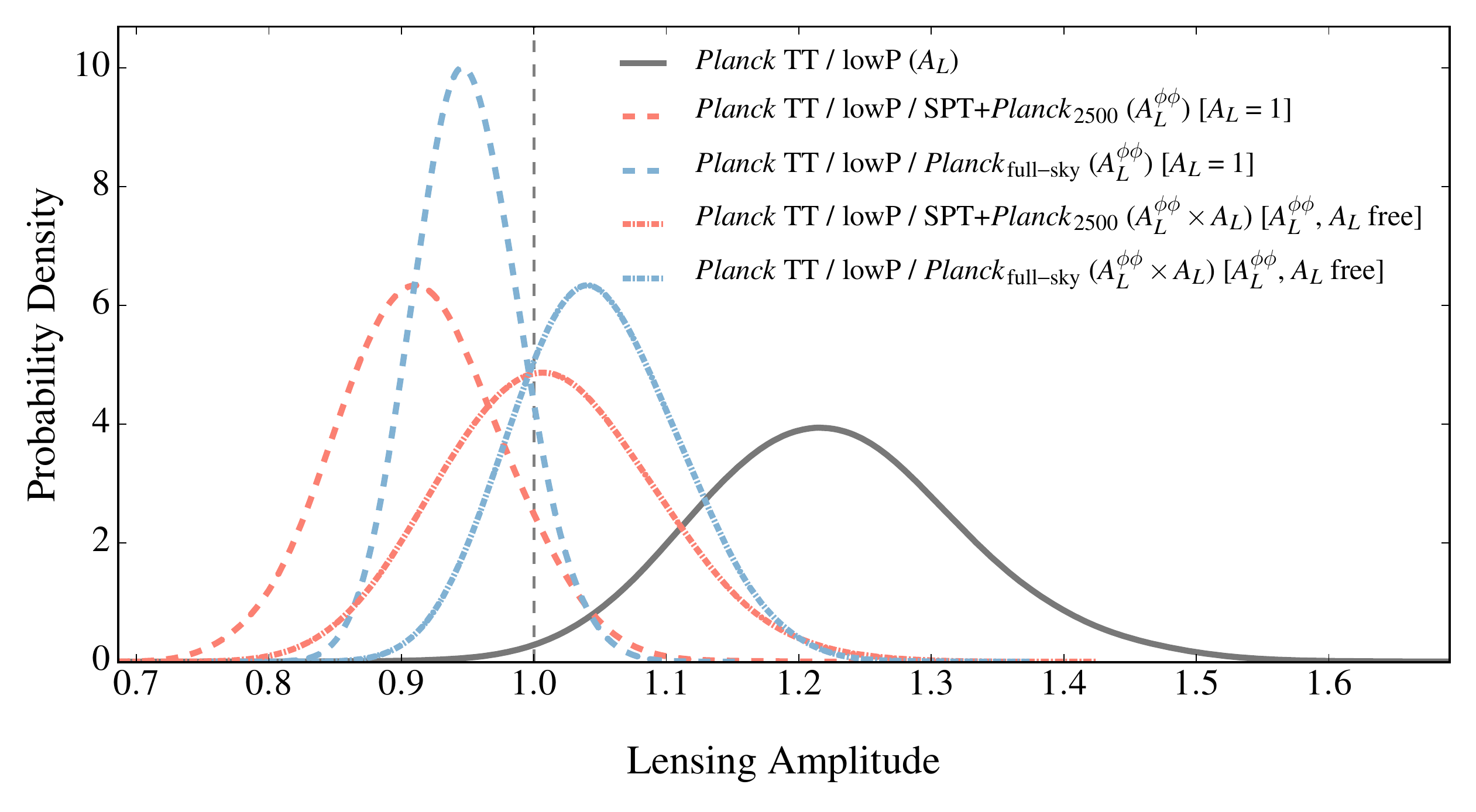}
\caption{Lensing amplitude constraints. The solid line shows the marginalized posterior
distribution of \alens\ from a fit using only \planck\ primary CMB power spectrum data. 
This is a measure of the level of smearing of the acoustic peaks relative to the prediction 
for \lcdm\ cosmological parameters from the \planck\ power spectrum ignoring the peak
smearing information.
Dashed lines show posterior distributions for \aphiphi\ from fits to \planck\ power spectrum
and full-sky \planck\ or 2500 \sqdeg \sptplanck\ lensing reconstruction data in which only the lensing
reconstruction power spectrum is allowed to vary, and the peak smearing is
constrained to be that expected for \lcdm. This is a measure of the lensing reconstruction
amplitude relative to the prediction for \lcdm\ cosmological parameters from the \planck\ 
power spectrum including the peak smearing information. Finally, the dot-dashed lines 
show posterior distributions for $\aphiphi \times \alens$, i.e., the lensing reconstruction
amplitude relative to predictions for \lcdm\ cosmological parameters from the \planck\ 
power spectrum ignoring the peak smearing information, from a fit in which 
\aphiphi\ and \alens\ are allowed to vary. }

\label{fig:alens}
\end{figure*}

Gravitational lensing of the CMB leads to a small amount of smearing
of the acoustic oscillations in the primary fluctuations, an effect
that has been well-measured \citep{das11b, keisler11}.  
The primary CMB fluctuations as observed by \planck\ show
weak evidence for a slightly elevated amount of lensing-like smearing
of the acoustic peaks, although the lensing power directly measured
by \planck\ shows no such excess \citep{planck15-15}. Using the same
effect in SPT temperature \citep{story13} and polarization
\citep{henning17} power spectra, there was no evidence for such an excess
of peak smoothing, with a modest ($\sim 1\sigma$) preference for 
less lensing than expected.

The expected amount of lensing depends on the somewhat uncertain
cosmological parameters.  To explore this, we marginalize over
cosmological parameters and 
use new parameters to artificially scale the amount of lensing:
\alens\ scales the lensing power spectrum in both the 
lens reconstruction power and in the smearing of the acoustic peaks,
and \aphiphi\ scales only the amplitude of the CMB lensing
reconstruction power spectrum. 
This parameterization ensures that the \lcdm\ parameters that control
the predicted degree of lensing (such as \omb and $\sigma_8$) are
determined without considering the measured amount of peak
smearing or mode coupling, and that these measurements are reflected
entirely in \alens\ and \aphiphi.

As these parameters are defined, the
\lcdm\ prediction for the reconstructed lensing power spectrum gets
multiplied by both \alens\ and \aphiphi. 
Therefore, the combination $\aphiphi \times \alens$ represents the 
amplitude for the lensing power relative to the \lcdm\ prediction
when the cosmological parameter fits are not sensitive to the 
observed amount of peak smearing.

With \alens\ fixed to unity
the known preference in the \planck\ primary data for $\alens > 1$
will instead drive a preference for models with higher
intrinsic lensing amplitudes, leading to a preference for lower values
of \aphiphi\ when compared with lensing reconstruction measurements that
are otherwise consistent with \lcdm. When \alens\ is free, the 
peak-smoothing preference for \alens$>1$ increases the predicted 
lensing reconstruction power and therefore causes a 
lower \aphiphi\ for a given model compared with the
lensing power spectrum measurement. The combination 
$\alens \times \aphiphi$ thus gives the amplitude of the lensing power spectrum 
compared to \planck-allowed \lcdm\ predictions when the peak smoothing
effect is not reflected in the \planck\ constraints.

Posterior distributions for \alens, \aphiphi, and $\aphiphi \times \alens$ from chains using
combinations of \planck\ primary CMB data, the \planck\ lensing power spectrum, 
and the lensing power spectrum in this work are shown in Figure~\ref{fig:alens}.
For models with \alens=1 the measured \sptplanck\ lensing power spectrum is somewhat low, 
with $\aphiphi=0.91 \pm 0.06$.
The CMB lensing reconstruction power spectrum
measurements show no evidence for an anomalous amount of lensing
relative to the amount predicted from the best-fit \lcdm\ parameters determined 
in the primary CMB data when the peak smearing effect has been marginalized over.
Using \sptplanck\ data, we find $\aphiphi \times \alens =1.01 \pm 0.08$ relative to 
the predicted level of lensing for \lcdm\ marginalized over \alens; 
using the full-sky \planck\ full-sky lensing reconstruction, the result is only slightly higher,
$\aphiphi \times \alens =1.05 \pm 0.06$.
The peak smearing in the 
\planck\ primary CMB power spectra, meanwhile, indicates mild evidence for
enhanced lensing, with \alens=$1.22 \pm 0.10$.

\subsection{Spatial curvature}

\begin{figure}[t!]
\centering
\includegraphics[width=1.1\columnwidth]{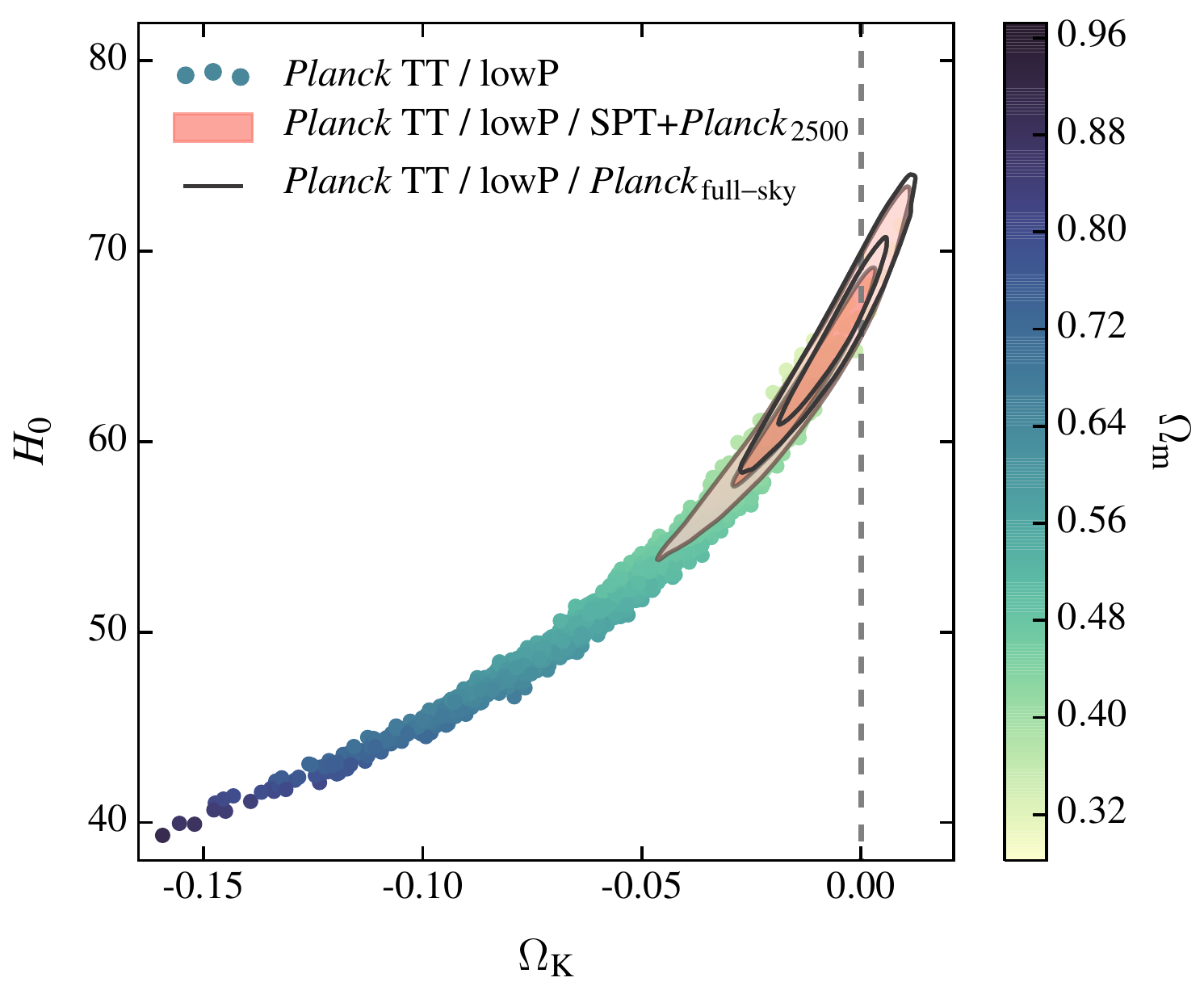}
\caption{Constraints on curvature for \planck\ primary CMB power spectra alone
(colored points), and adding either full-sky \planck\ lensing (black contours)
or 2500 \sqdeg \sptplanck\ lensing data (green). Points from the no-lensing chain are color-coded
by the matter density.}
\label{fig:omk}
\end{figure}

Inflationary models predict that the universe should be close to spatially 
flat, and the combination of observations of the primary CMB, supernovae Ia, 
baryon acoustic oscillations, and local Hubble constant measurements 
show that spatial curvature is not large \citep[e.g.,][]{komatsu11}.
Constraints from the primary CMB have a geometrical degeneracy that allows spatial curvature to
be increased while the Hubble constant is adjusted to keep the angular diameter distance
to last scattering fixed; as a result, CMB measurements have historically relied
on Hubble constant priors or external measurements to constrain curvature. 
As a probe of the local universe, lensing partially lifts  this degeneracy \citep{sherwin11}.
Figure~\ref{fig:omk} demonstrates this degeneracy-breaking by adding 2500 \sqdeg\ 
\sptplanck\ or  
\planck\ full-sky lensing reconstruction information to the \planck\ primary CMB 
measurements. The constraint on spatial curvature from adding \sptplanck\ lensing
information to \planck\ primary CMB is $\Omega_k = -0.012^{+0.021}_{-0.023}$ at 95\% confidence. 

\subsection{Massive neutrinos}

CMB lensing, as a measurement of
the amplitude of clustering at intermediate redshifts, is a potentially 
powerful probe of neutrino masses \citep{smith06b,abazajian15b}. 
Neutrino oscillation experiments
have precisely measured the differences in the squares of the masses
between the neutrino eigenstates, but the absolute masses have not
been measured. Laboratory limits constrain the mass of the electron
neutrino, but the strongest constraints on absolute neutrino
masses currently come from cosmology. 
Having a substantial amount of
the energy density in the form of massive neutrinos leads to
a suppression of structure on small scales in the matter power spectrum.
The \planck\ primary CMB measurements
limit the sum of the masses to be  $\mnu < 0.72$ eV at 95\% confidence \citep{planck15-13}. 
This constraint is
strongly driven by the measurement of lensing through the smearing of peaks in the CMB power spectra. 

\begin{figure}
\centering
\includegraphics[width=0.875\columnwidth]{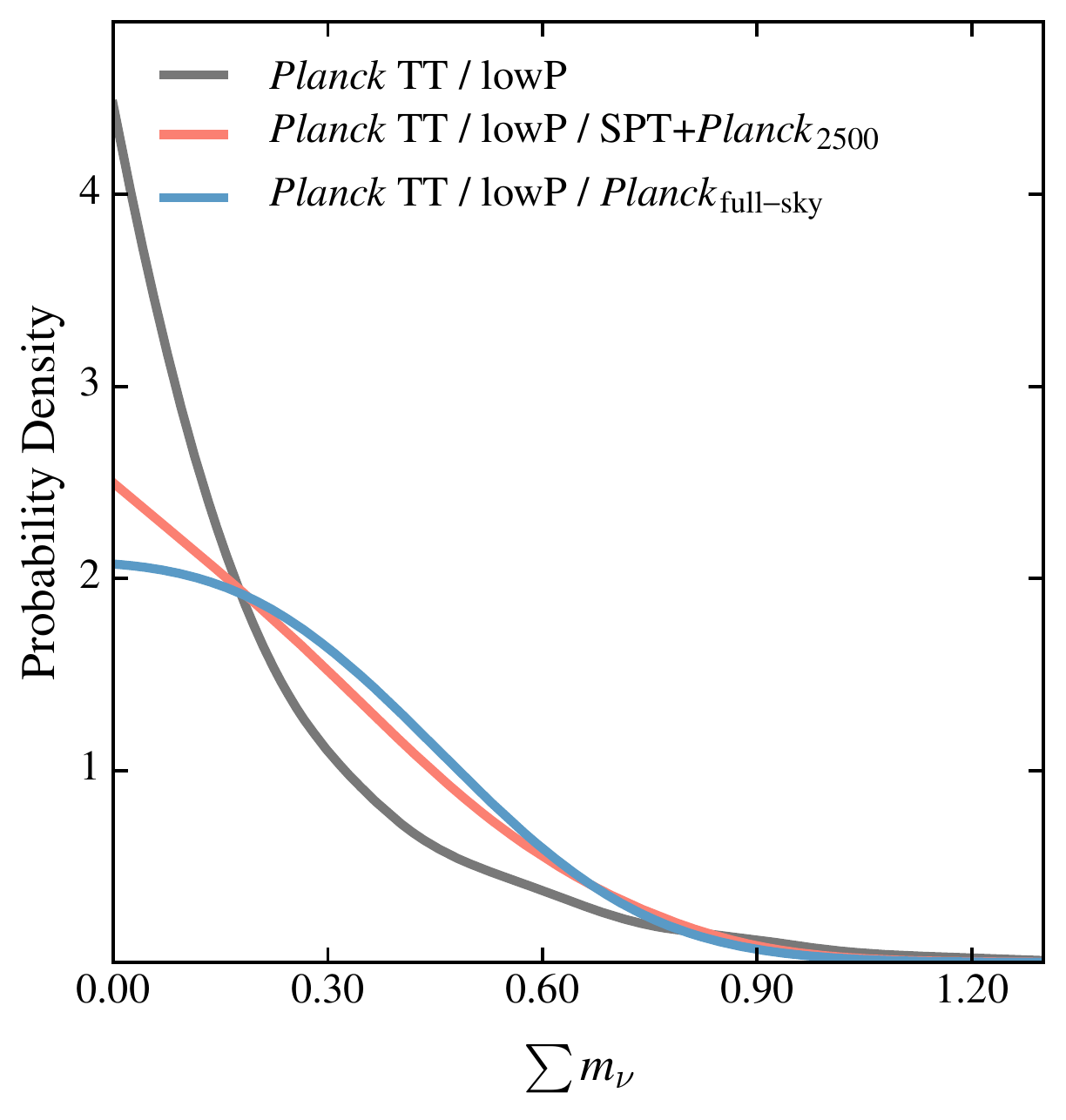}
\caption{Limits on neutrino masses, showing \planck\ power spectrum information
on its own (grey), and 
adding either \planck\ lensing data (blue) or \sptplanck\ information (red)}
\label{fig:mnu}
\end{figure}

As was seen in \citet{planck15-13}, Figure~\ref{fig:mnu} shows that adding information from the lensing 
reconstruction power spectrum reduces the \mnu\ posterior value at zero, but
the lensing reconstruction data also rule out large values of \mnu, with the 
combined result being a similar 95\% upper limit.
Using \sptplanck, the upper limit on neutrino masses is
$\mnu<0.70$ eV at 95\% confidence, compared to 
$\mnu<0.68$ eV for adding \planck\ full-sky lensing reconstruction data.

\section{Discussion}

The \sptplanck\ data are not quite as constraining as the \planck-only lensing
constraints; while the signal-to-noise per pixel of the O17
lensing map is substantially higher, the combined map covers only 2500 \sqdegnospace. 
As discussed in O17, the statistical
precision of the combined lensing map is dominated by the SPT data.  
This measurement is therefore a nearly
independent check on the \planck\ lensing measurement. 

The \sptplanck\ lensing measurements and resulting cosmological
constraints are in close agreement with the 
full-sky lensing results of the \planck\ experiment. 
For example, the \sptplanck\ lensing measurements are in excellent agreement with a 
spatially flat universe, as predicted by inflationary models, with 
$\Omega_k = -0.012^{+0.21}_{-0.023}$ at 95\% confidence, while a constraint on
local structure from \sptplanck\ is $\sigom=0.601 \pm 0.023$. Using
\planck\ lensing instead yields $\Omega_k=-0.005^{+0.016}_{-0.017}$
and $\sigom=0.607 \pm 0.015$.
These new measurements are nearly statistically independent of the \planck-only results,
so the agreement between the datasets is informative.

This trend is also true for slight tensions that exist in \planck\ between the amount
of lensing inferred from peak smearing and the direct reconstructions from the higher-order 
statistics. Measurements of the lensing amplitude from \sptplanck\ are in excellent
agreement with the lensing amplitude inferred from the \planck\ higher-order statistics,
and in slight tension with that inferred from CMB peak smearing. When marginalizing over
the peak smearing effect, the \sptplanck\ data are in close agreement with the
expected amount of gravitational lensing otherwise predicted by the observed
CMB fluctuations.

The amount of lensing seen in \sptplanck\ is also broadly consistent with both the
amplitude inferred from low-redshift galaxy lensing studies and the amplitude of
structure inferred from the \planck\ primary CMB measurements, as was also the
case with \planck-only lensing constraints. More precise CMB lensing measurements will
be required to further investigate possible tensions between low-redshift and high-redshift
determinations of the amplitude of structure.

\section{Acknowledgments}

	G. S. wishes to thank Elisa G. M. Ferreira, Joachim Harnois-D\'eraps and Alexander van Engelen for useful discussions and Jack Holder for 
	digitization of the foreground model.
	We acknowledge the use of Alexander van Engelen's implementation of the analytical $N^{(1)}_{L}$ bias in the flat-sky approximation. 
    	The South Pole Telescope program is supported by the National Science Foundation through grant PLR-1248097. Partial support is also provided by the NSF Physics Frontier Center grant PHY-0114422 to the Kavli Institute of Cosmological Physics at the University of Chicago, the Kavli Foundation, and the Gordon and Betty Moore Foundation through Grant GBMF\#947 to the University of Chicago. 
This work has made use of computations performed on Guillimin, managed by
Calcul Quebec and Compute Canada (funded by CFI, MESI, and FRQNT), 
and the Blue Waters sustained-petascale
computing project (supported by NSF awards OCI-0725070 and ACI-1238993 and
the state of Illinois).
The McGill authors acknowledge funding from the Natural Sciences and Engineering Research Council of Canada, Canadian Institute for Advanced Research, and Canada Research Chairs program.
	G. S. acknowledges support from the Fonds de recherche du Qu\'ebec - Nature et technologies.
BB has been supported by the Fermi Research Alliance, LLC under Contract No. DE-AC02-07CH11359 with the U.S. Department of Energy, Office of Science, Office of High Energy Physics.
CR acknowledges support from Australian Research CouncilÕs Discovery Projects scheme (DP150103208).

\bibliography{../../../BIBTEX/spt}

\end{document}

%% file: spt_authorlist_v1.tex

\newcommand{\McGill}{Department of Physics and McGill Space Institute, McGill University, Montreal, Quebec H3A 2T8, Canada}
\newcommand{\KIPAC}{Kavli Institute for Particle Astrophysics and Cosmology, Stanford University, 452 Lomita Mall, Stanford, CA 94305}
\newcommand{\Stanford}{Dept. of Physics, Stanford University, 382 Via Pueblo Mall, Stanford, CA 94305}
\newcommand{\Davis}{Department of Physics, University of California, Davis, CA, USA 95616}
\newcommand{\Penn}{Center for Particle Cosmology, Department of Physics and Astronomy, University of Pennsylvania, Philadelphia, PA,  USA 19104} 
\newcommand{\KICPChicago}{Kavli Institute for Cosmological Physics, University of Chicago, Chicago, IL, USA 60637}
\newcommand{\PhysicsUChicago}{Department of Physics, University of Chicago, Chicago, IL, USA 60637}
\newcommand{\AAUChicago}{Department of Astronomy and Astrophysics, University of Chicago, Chicago, IL, USA 60637}
\newcommand{\FNAL}{Fermi National Accelerator Laboratory, MS209, P.O. Box 500, Batavia, IL 60510}
\newcommand{\ArgonneHEP}{High Energy Physics Division, Argonne National Laboratory, Argonne, IL, USA 60439}
\newcommand{\EFIChicago}{Enrico Fermi Institute, University of Chicago, Chicago, IL, USA 60637}
\newcommand{\SLAC}{SLAC National Accelerator Laboratory, 2575 Sand Hill Road, Menlo Park, CA 94025}
\newcommand{\Caltech}{California Institute of Technology, Pasadena, CA, USA 91125}
\newcommand{\Berkeley}{Department of Physics, University of California, Berkeley, CA, USA 94720}
\newcommand{\Cifar}{Canadian Institute for Advanced Research, CIFAR Program in Cosmology and Gravity, Toronto, ON, M5G 1Z8, Canada}
\newcommand{\Colorado}{Center for Astrophysics and Space Astronomy, Department of Astrophysical and Planetary Sciences, University of Colorado, Boulder, CO, 80309}
\newcommand{\ESO}{European Southern Observatory, Karl-Schwarzschild-Stra{\ss}e 2, 85748 Garching, Germany}
\newcommand{\Colphys}{Department of Physics, University of Colorado, Boulder, CO, 80309}
\newcommand{\Illast}{Astronomy Department, University of Illinois at Urbana-Champaign, 1002 W. Green Street, Urbana, IL 61801, USA}
\newcommand{\Illphys}{Department of Physics, University of Illinois Urbana-Champaign, 1110 W. Green Street, Urbana, IL 61801, USA}
\newcommand{\UChicago}{University of Chicago, Chicago, IL, USA 60637}
\newcommand{\LBNL}{Physics Division, Lawrence Berkeley National Laboratory, Berkeley, CA, USA 94720}
\newcommand{\Michigan}{Department of Physics, University of Michigan, Ann  Arbor, MI, USA 48109}
\newcommand{\Munich}{Faculty of Physics, Ludwig-Maximilians-Universit\"{a}t, 81679 M\"{u}nchen, Germany}
\newcommand{\ExcellenceCluster}{Excellence Cluster Universe, 85748 Garching, Germany}
\newcommand{\MPE}{Max-Planck-Institut f\"{u}r extraterrestrische Physik, 85748 Garching, Germany}
\newcommand{\Dunlap}{Dunlap Institute for Astronomy \& Astrophysics, University of Toronto, 50 St George St, Toronto, ON, M5S 3H4, Canada}
\newcommand{\Minnesota}{Department of Physics, University of Minnesota, Minneapolis, MN, USA 55455}
\newcommand{\Melbourne}{School of Physics, University of Melbourne, Parkville, VIC 3010, Australia}
\newcommand{\CaseWestern}{Physics Department, Center for Education and Research in Cosmology and Astrophysics, Case Western Reserve University,Cleveland, OH, USA 44106}
\newcommand{\ArtInstChicago}{Liberal Arts Department, School of the Art Institute of Chicago, Chicago, IL, USA 60603}
\newcommand{\JPL}{Jet Propulsion Laboratory, California Institute of Technology, Pasadena, CA 91109, USA}
\newcommand{\CfA}{Harvard-Smithsonian Center for Astrophysics, Cambridge, MA, USA 02138}
\newcommand{\UToronto}{Department of Astronomy \& Astrophysics, University of Toronto, 50 St George St, Toronto, ON, M5S 3H4, Canada}
\newcommand{\BCCP}{Berkeley Center for Cosmological Physics, Department of Physics, University of California, and Lawrence Berkeley National Labs, Berkeley, CA, USA 94720}



\author{G.~Simard}
\affiliation{\McGill}

\author{Y.~Omori}
\affiliation{\McGill}
\affiliation{\KIPAC}
\affiliation{\Stanford}

\author{K.~Aylor}
\affiliation{\Davis}

\author{E.~J.~Baxter}
\affiliation{\Penn}
\affiliation{\KICPChicago}
\affiliation{\AAUChicago}  

\author{B.~A.~Benson}
\affiliation{\FNAL}
\affiliation{\KICPChicago}
\affiliation{\AAUChicago}

\author{L.~E.~Bleem}
\affiliation{\ArgonneHEP} 
\affiliation{\KICPChicago}

\author{J.~E.~Carlstrom}
\affiliation{\KICPChicago}
\affiliation{\PhysicsUChicago}
\affiliation{\ArgonneHEP}
\affiliation{\AAUChicago}
\affiliation{\EFIChicago}

\author{ C.~L.~Chang}
\affiliation{\ArgonneHEP}
\affiliation{\KICPChicago}
\affiliation{\AAUChicago}

\author{ H-M.~Cho}
\affiliation{\SLAC}

\author{ R.~Chown}
\affiliation{\McGill}

\author{ T.~M.~Crawford}
\affiliation{\KICPChicago}
\affiliation{\AAUChicago}

\author{ A.~T.~Crites}
\affiliation{\KICPChicago}
\affiliation{\AAUChicago}
\affiliation{\Caltech}

\author{ T.~de~Haan}
\affiliation{\McGill}
\affiliation{\Berkeley}

\author{ M.~A.~Dobbs}
\affiliation{\McGill}
\affiliation{\Cifar}

\author{ W.~B.~Everett}
\affiliation{\Colorado}

\author{ E.~M.~George}
\affiliation{\Berkeley}
\affiliation{\ESO}

\author{ N.~W.~Halverson}
\affiliation{\Colorado}
\affiliation{\Colphys}

\author{ N.~L.~Harrington}
\affiliation{\Berkeley}

\author{ J.~W.~Henning}
\affiliation{\ArgonneHEP}
\affiliation{\KICPChicago}

\author{ G.~P.~Holder}
\affiliation{\McGill}
\affiliation{\Cifar}
\affiliation{\Illast}
\affiliation{\Illphys}

\author{ Z.~Hou}
\affiliation{\KICPChicago}
\affiliation{\AAUChicago}

\author{ W.~L.~Holzapfel}
\affiliation{\Berkeley}

\author{ J.~D.~Hrubes}
\affiliation{\UChicago}

\author{ L.~Knox}
\affiliation{\Davis}

\author{ A.~T.~Lee}
\affiliation{\Berkeley}
\affiliation{\LBNL}

\author{ E.~M.~Leitch}
\affiliation{\KICPChicago}
\affiliation{\AAUChicago}

\author{ D.~Luong-Van}
\affiliation{\UChicago}

\author{ A.~Manzotti}
\affiliation{\KICPChicago} 
\affiliation{\AAUChicago} 

\author{ J.~J.~McMahon}
\affiliation{\Michigan}

\author{ S.~S.~Meyer}
\affiliation{\KICPChicago}
\affiliation{\AAUChicago}
\affiliation{\EFIChicago}
\affiliation{\PhysicsUChicago}

\author{ L.~M.~Mocanu}
\affiliation{\KICPChicago}
\affiliation{\AAUChicago}

\author{ J.~J.~Mohr}
\affiliation{\Munich}
\affiliation{\ExcellenceCluster}
\affiliation{\MPE}

\author{ T.~Natoli}
\affiliation{\KICPChicago}
\affiliation{\PhysicsUChicago}
\affiliation{\Dunlap}

\author{ S.~Padin}
\affiliation{\KICPChicago}
\affiliation{\AAUChicago}

\author{ C.~Pryke}
\affiliation{\Minnesota}

\author{ C.~L.~Reichardt}
\affiliation{\Berkeley}
\affiliation{\Melbourne}

\author{ J.~E.~Ruhl}
\affiliation{\CaseWestern}

\author{ J.~T.~Sayre}
\affiliation{\CaseWestern}
\affiliation{\Colorado}

\author{ K.~K.~Schaffer}
\affiliation{\KICPChicago}
\affiliation{\EFIChicago}
\affiliation{\ArtInstChicago}

\author{ E.~Shirokoff}
\affiliation{\Berkeley} 
\affiliation{\KICPChicago} 
\affiliation{\AAUChicago} 

\author{ Z.~Staniszewski}
\affiliation{\CaseWestern}
\affiliation{\JPL}

\author{ A.~A.~Stark}
\affiliation{\CfA}

\author{ K.~T.~Story}
\affiliation{\KICPChicago}
\affiliation{\PhysicsUChicago}
\affiliation{\KIPAC}
\affiliation{\Stanford}

\author{ K.~Vanderlinde}
\affiliation{\Dunlap}
\affiliation{\UToronto}

\author{J.~D.~Vieira}
\affiliation{\Illast} 
\affiliation{\Illphys} 

\author{R.~Williamson}
\affiliation{\KICPChicago} 
\affiliation{\AAUChicago} 

\author{W.~L.~K.~Wu}
\affiliation{\KICPChicago}